\documentclass[12pt]{article}
\usepackage{graphicx}
\begin{document}

\begin{center}
{\bf \Large The mixing in the solar core and the neutrino fluxes}
\vskip 0.2in

  Anatoly Kopylov and Valery Petukhov \\
{\it Institute of Nuclear Research of Russian Academy of Sciences\\
117312 Moscow, Prospect of 60th Anniversary of October Revolution
7A}
\vskip 0.2in

\end{center}

\footnotetext{Corresponding author: Kopylov A.V., Institute for
Nuclear Research of Russian Academy of Sciences, Prospect of 60th
Anniversary of October Revolution 7A, 117312, Moscow, Russia;
telephone +7(495)8510961, e-mail: beril@inr.ru }

\begin{abstract}
The question is addressed whether it is possible to introduce a
mixing in the solar model not in a contradiction with the present
data of helioseismology. As a new thing it is shown that there is
indeed a spherical geometry with mixing for which the sound speed
profile would be consistent with the helioseismic data. The effect
of such mixing in a spherical shell in the central zone of the Sun
would be the substantial increase of the neutrino flux from $^{13}N$
while all other neutrino fluxes will stay practically unchanged. The
implications for future experiments are discussed.
\end{abstract}

\underline{solar neutrinos; neutrino experiments}
\vskip 0.2in

{\bf Introduction.}

Neutrinos are generated in the Sun in the reactions of pp-chain and
in CNO cycle. The latter contributes only about 1\% to the total
solar energy, but neutrinos generated in this cycle may carry very
substantial information about the processes deep in the interior of
the Sun. The fluxes of CNO neutrinos depend primarily on the
abundances of C and N in the solar core where CNO neutrinos are
generated. This was stated by G.Zatsepin and V.Kuzmin \cite{1} and
also by J.Bahcall \cite{2} on the eve of the solar neutrino
research, particularly, formulating the motivation for a lithium
experiment on solar neutrinos. Since that time a tremendous success
in the study of solar neutrinos has been achieved
\cite{3}--\cite{18}. The high temperature dependence of the
reactions of CNO cycle is no more our prime concern, because the
flux of $^8B$ neutrinos has been proven to be very close (within
1$\sigma$) to the prediction of the solar model \cite{19}. Due to
the correlation of the fluxes of $^8B$ and CNO neutrinos, as it was
first emphasized in \cite{20} and then was presented in more details
in \cite{21}, the uncertainty in the temperature should not be the
greatest factor varying the fluxes of CNO neutrinos. But what about
the abundances of carbon and nitrogen, can this be a major factor?
The data on photospheric abundances of the Sun indicate the lower
composition of metals than the standard model suggests, but this
conflicts with the results of helioseismology \cite{22}--\cite{31}.
Thus, we have by now two alternatives: an old model with high
metallicity which is in a very good agreement with helioseismology
and a new model with low metallicity which is at odds with
helioseismology \cite{32}. The fluxes of neutrinos predicted by
these models \cite{31, 33} are quite different, especially what
concerns CNO neutrinos. There's another aspect of this question. As
one can see from Figure 1, the profiles of the abundances of these
elements across the Sun are also very different. The abundance of
$^{14}N$ varies by a factor of 4 while the abundance of $^{12}C$
varies by more than two orders of magnitude. This may have a
dramatic influence on the fluxes of solar neutrinos. In our previous
paper \cite{34} it was shown that the flux $F_{13}$ of $^{13}N$
neutrinos can be substantially (more than 30\%) increased by local
mixing in a spherical shell of the central region of the Sun.  The
fluxes $F_7$ of $^7Be$ and $F_{15}$ of $^{15}O$ neutrinos experience
by this mixing only a tiny change on the level of a few percent.
Remarkably that the mean molecular weight averaged for the central
zone of the Sun, $\mu_c$, has been changed by this mixing only on a
level of less than 0.5\% what, in fact, is in agreement with the
most accurately determined value fixed at present by
helioseismology. However, the local changes of the mean molecular
weight within the area of mixing can reach the value of a few
percent. Then it may result in a substantial deviation of the sound
speed profile in this region. The sound speed is connected with
$\mu$ by the expression:

\begin{equation} \label{eq1}
c=\sqrt{\frac{\gamma kT}{\mu } }
\end{equation}

where $\gamma \approx 5/3$ -- the adiabatic constant, \textit{k} --
the Boltsman constant. So it is quite interesting to see whether
there could be a mixing which would not change the parameters fixed
by helioseismology and the fluxes of solar neutrinos generated in
pp-chain precisely determined by experiments \cite{2}--\cite{8} and
at the same time will produce substantial increase of $F_{13}$ while
the flux $F_{15}$ will stay practically unchanged. Here it is worth
to note that for the solar models with high (GS98) \cite{21} and low
(AGS09)\cite{35} abundance of CNO elements the difference of the
fluxes $F_{13}$ and $F_{15}$ is about 30\% while for the ratio of
these fluxes the difference is less than 10\% (see, for example,
Table 2 of \cite{33}). Thus the ratio $F_{13}/F_{15}$ appears to be
a sensitive indicator of mixing in the solar core what can be
exploited in the future experiments on solar neutrinos.

\vskip 0.2in
{\bf Simulation of mixing.}

The calculation has been performed with the following limitations
for mixing: the mass, density and temperature of the spherical shell
is not changed by mixing. The process is very slow and there is time
for these parameters to be restored.  The resulted abundance of
elements in the mixed shell is determined by several parameters: the
position and thickness of the shell, the intensity of mixing, its
duration etc. No attempts has been made to describe the physics of
mixing i.e. to determine the driving force, the magneto
hydrodynamics  etc. Only one illustrative case of mixing has been
considered with the aim to see what can be the effect of such
mixing, provided this process is realized. It was assumed that
mixing in the center of a shell is more intensive than on its
periphery with a rather smooth transient function. In this case the
profile for the abundances of different elements can look how it is
presented on Figure 1. The initial profile was taken from the model
AGS05 \cite{36}.

\begin{figure}[!ht]
\centering
\includegraphics[width=4in]{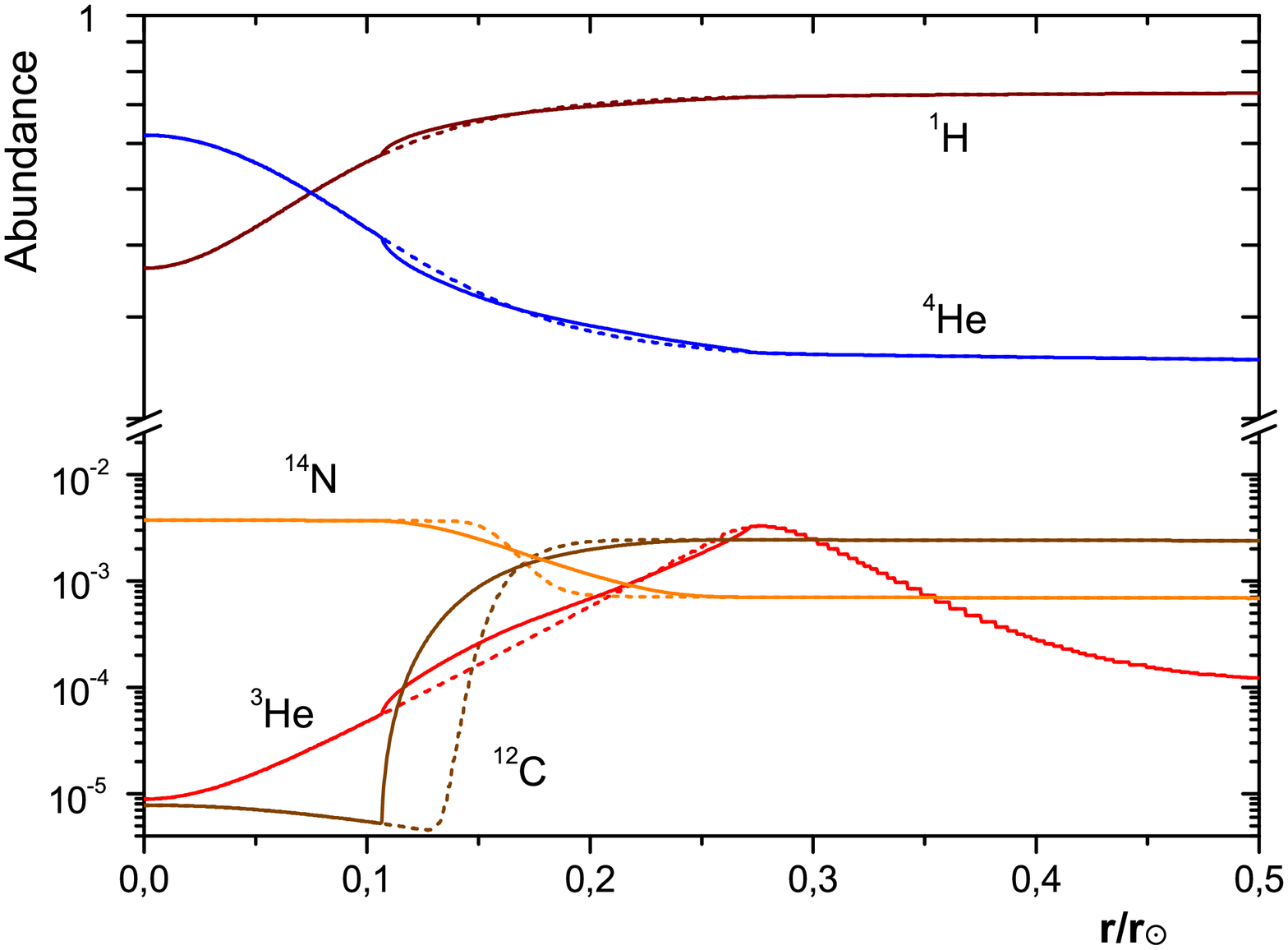}
\caption{The profile for the abundance of $^1H$, $^3He$, $^4He$,
$^{12}C$ and $^{14}N$ without (dashed line) and with (solid line)
mixing.}
\end{figure}

The sound speed profile by this mixing will be changed differently
for two sub shells relatively to the one fixed by helioseismology as
it is shown on Figure 2.

\begin{figure}[!ht]
\centering
\includegraphics[width=4in]{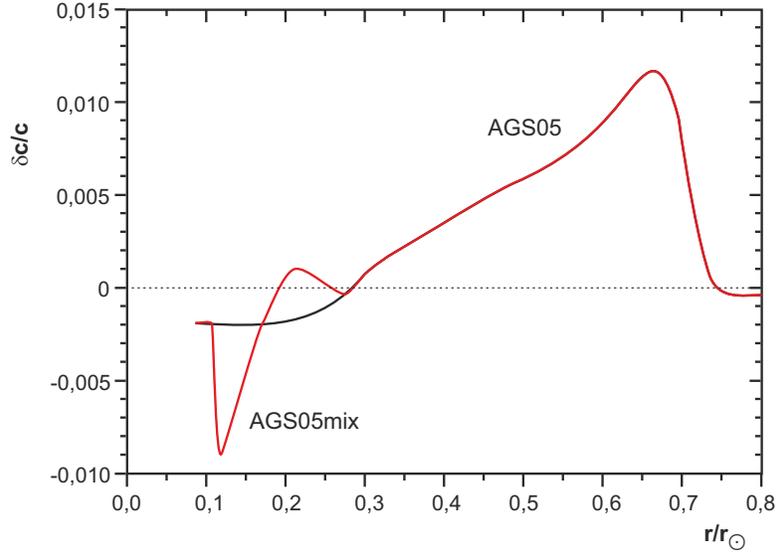}
\caption{The differences between the helioseismic and predicted
sound speed as a function of depth without \cite{32} (black line)
and with (red line) mixing.}
\end{figure}

As one can see from this figure there's improved agreement of the
mixed model with the helioseismic data for the external part of the
mixed shell. The maximal deviation of the calculated value $\delta
c/c$ from helioseismic data $<$ 1\% at the depth 0.12$R_{\odot}$.
The change of the mean molecular weight of the solar core is $<$
0.2\%. The flux $F_{13}$ is determined by the reaction rate
$\alpha_{112}$ of the reaction $^{12}C + p \to {} ^{13}N + \gamma$
and the abundances $X_1(r)$ and $X_{12}(r)$ of isotopes $^1H$ and
$^{12}C$ along the solar radius. The nucleus $^{13}N$ produced in
this reaction quickly decays generating $^{13}N$-neutrino and
positron.

\begin{equation} \label{eq2}
F_{13}(t)=4\pi N_{A} R_{\odot }^{3} \int _{0}^{1}r^{2} \rho (r)X_{1}
(r,t)\frac{X_{12} (r,t)}{12}  \alpha _{112} (r)dr
\end{equation}

The similar expressions can be written for the fluxes $F_{15}$ and
$F_7$ of $^{15}O$- and $^7Be$-neutrinos. The results of the
calculation are presented in Table 1. The fluxes without mixing were
obtained in a toy model with the parameters close to AGS05. As one
can see from Table 1 only the flux $F_{13}$ experiences a
substantial increase ($\sim$ 90\%) in the model with mixing. The
change is high enough to be observed by experiment.

\small
\begin{table}[!ht]
\centering \caption{Solar neutrino fluxes with and without mixing,
in units of $10^9(F_7)$ and $10^8(F_{13}, F_{15})$ $cm^{-2}s^{-1}$}
\begin{tabular}{p{2cm}p{2cm}p{2cm}p{3cm}}
\hline
Flux & No mix flux & Mix flux & Difference, \% \\
\hline $F_7$ & 4,43 & 4,65 & 4,8 \\
\hline $F_{13}$ & 1,99 & 3,77E & 89,8 \\
\hline $F_{15}$ & 1,58 & 1,57 & -0,14 \\
\hline
\end{tabular}
\end{table}

\vskip 0.2in
{\bf Conclusions.}

The abundance of $^{12}C$ along the radius of the Sun has a peculiar
feature at the depth of 0.15 of solar radius where it experiences a
sharp increase by more than two orders of magnitude. If there is a
mixing, even a very mild one, in the shell at this depth, the result
can be a substantial increase of the flux of $^{13}N$ neutrinos. The
problem is that present data of helioseismology, especially the
sound speed profile, restrict severely the possibility of such
mixing. As a new thing, here it is shown that there is indeed a
spherical geometry with mixing for which the sound speed profile
would be consistent with the helioseismic data. In case of such
mixing it shown also that all other neutrino fluxes are left
unchanged, at most they will vary on a level of a few percent. The
difference of the fluxes $F_{13}$ and $F_{15}$ for models with high
and low abundance of CNO elements reach a substantial value of about
30\% while the difference for the ratio of these fluxes is $<$ 10\%.
It appears to be that the ratio $F_{13}/F_{15}$ is a parameter very
sensitive to mixing in the solar core. Only future experiments may
clarify the question whether the mixing exists or not. The
experiments on solar neutrinos are in a mature phase now. The
sensitivity and precision increased remarkably during last 30 years
\cite{37}--\cite{39}. The new projects LENA \cite{40}, SNO+
\cite{41} etc. have increased sensitivity to CNO neutrinos. Here we
would like to draw attention to the discovering potential of finding
the anomalous high flux $F_{13}$ and may be the ratio of $F_{13}$ to
$F_{15}$. This would indicate a mixing in the core of the Sun and
other solar type stars of main sequence.

\vskip 0.2in
{\bf Acknowledgements.}

We warmly acknowledge funding from the Programs of support of
leading schools of Russia (grant \#871.2012.2.)

\end{document}